\documentclass{iau}
\usepackage{amsmath}
\usepackage{amssymb}
\usepackage{graphicx}
\usepackage{natbib}

\title[Radio SED of highly star-forming galaxies] 
{Average radio spectral energy distribution of highly star-forming galaxies}

\author[K. Tisani\'{c} et al.]  
{
     K. Tisani\'{c}$^1$
    \and V. Smol\v{c}i\'{c}$^1$  
    \and J. Delhaize$^1$
    \and M. Novak$^1$ 
    \and H. Intema$^2$ 
    \and I. Delvecchio$^1$
    \and E. Schinnerer$^3$
    \and G. Zamorani$^4$
    \affiliation{
    $^1$Department of Physics,
    Faculty of Science,
    University of Zagreb,
    Bijeni\v{c}ka cesta 32,
    10000  Zagreb, Croatia,  email: {\tt ktisanic@phy.hr} \\[\affilskip] $^2$Leiden Observatory,
    Leiden University,
    Niels Bohrweg 2,
    2333 CA Leiden,
    The Netherlands\\[\affilskip] $^3$MPI for Astronomy, K\"onigstuhl 17, D-69117, Heidelberg, Germany\\[\affilskip] $^4$INAF-Osservatorio Astronomico di Bologna, via Gobetti 93/3, 40129, Bologna, Italy}}

\usepackage{url}
\pubyear{2018}
\volume{333}  
\jname{Peering towards Cosmic Dawn}
\editors{Vibor Jeli\'c \& Thijs van der Hulst, eds.}
\begin{document}
\newcommand{\GHz}{\,\mathrm{GHz}}
\newcommand{\MHz}{\,\mathrm{MHz}}
\newcommand{\kHz}{\,\mathrm{kHz}}
\newcommand{\km}{\,\mathrm{km}}
\newcommand{\hr}{\,\mathrm{hours}}
\newcommand{\yr}{\,\mathrm{yr}}
\newcommand{\Msun}{\,\mathrm{M_{\odot}}}

\newcommand{\delvip}{refId1}

\maketitle

\begin{abstract}
The infrared-radio correlation (IRRC) offers a way to assess star formation from radio emission. Multiple studies found the IRRC to decrease with increasing redshift. This may in part be due to the lack of knowledge about the possible radio spectral energy distributions (SEDs) of star-forming galaxies. We constrain the radio SED of a complete sample of highly star-forming galaxies ($SFR>100\Msun/\yr$) based on the VLA-COSMOS $1.4\GHz$ Joint and $3\GHz$ Large Project catalogs. We reduce  archival GMRT $325\MHz$ and $610\MHz$ observations, broadening the rest-frame frequency range to $0.3-15\GHz$. Employing survival analysis and fitting a double power law SED, we find that the slope steepens from a spectral index of $\alpha_1=0.51\pm 0.04$ below $4.5\GHz$ to $\alpha_2=0.98\pm0.07$ above $4.5\GHz.$ Our results suggest that the use of a K-correction assuming a single power-law radio SED for star forming galaxies is likely not the root cause of the IRRC trend.
\end{abstract}
\firstsection 
\section{Introduction}
Star formation rate measurements (SFRs) derived from the ultraviolet and optical emission are sensitive to dust \citep{Kennicutt98}, and infrared-derived SFRs are easy to understand only in the optically thick case \citep{Kennicutt98, Condon92}.
On the other hand, radio emission in star-forming galaxies below  $\sim30\GHz$  is dust-unbiased and  is thought to be  primarily produced by synchrotron radiation of ultra-relativistic electrons  accelerated by supernovae and free-free emission arising from HII regions \citep{Condon92}.

It has been found that the infrared and radio luminosities are bound by a tight correlation over many orders of magnitude \citep{Kruit71, Helou85}. This infrared-radio correlation,  defined by the so-called $q$ parameter\footnote{Defined as $q=\log\frac{L_{TIR}}{3.75\,\mathrm{THz}}-\log(L_{1.4\GHz})$, where $L_{TIR}$ is the total infrared luminosity ($8-1000\,\mathrm{\mu m}$) in units of $\mathrm{W}$ and $L_{1.4\GHz}$ is the luminosity density at $1.4\GHz$ in units of $\mathrm{W/Hz}$. Frequency of $3.75\mathrm{THz}$ was used to obtain a dimensionless quantity.}, provides the basis for radio luminosity as a star-formation tracer, as infrared luminosity is linked to SFR \citep{Kennicutt98b}. 
Recent stacking and survival analysis studies \citep{Ivison10,Magnelli15, Delhaize17, CalistroRivera17} find $q$ to be decreasing with increasing redshift.
\citet{Delhaize17}  point out that the computation of rest-frame radio luminosity via a K-correction using a simple, single power-law assumption of the star-forming galaxies' spectral energy distribution (SED) could possibly cause such a trend.
The exact shape of the radio SED of star-forming galaxies is usually assumed to be a superposition of the steep synchrotron spectrum, described by a power law, and a $\sim10\%$ contribution at $1.4\GHz$ of a flat free-free spectrum \citep{Condon92}. 
These results are supported by observations of  nearby galaxies at $3-30\,\mathrm{Mpc}$ ($z\approx 0.01$) with $SFR<10\Msun/\yr$ \citep{Tabatabaei17}.
For (ultra)luminous infrared galaxies ((U)LIRGs, $SFR>10 \Msun/\yr$), observations in the local universe find the SED to be steep around $10\GHz$, corresponding to a synchrotron spectrum, with  flattening below (above) $10\GHz$ due to free-free absorption (emission) \citep{Clemens08, Leroy11, Galvin17}. 

In this paper we determine the shape of the radio SED for a highly star-forming galaxy sample in the COSMOS field \citep{Scoville07}.
In Sects. 2, 3, and 4 we describe the data and sample, method used and present our results, respectively. We summarize in Sect. 5.
\section{Data and sample selection}
\begin{figure}[t]
\centering\includegraphics[height=4cm]{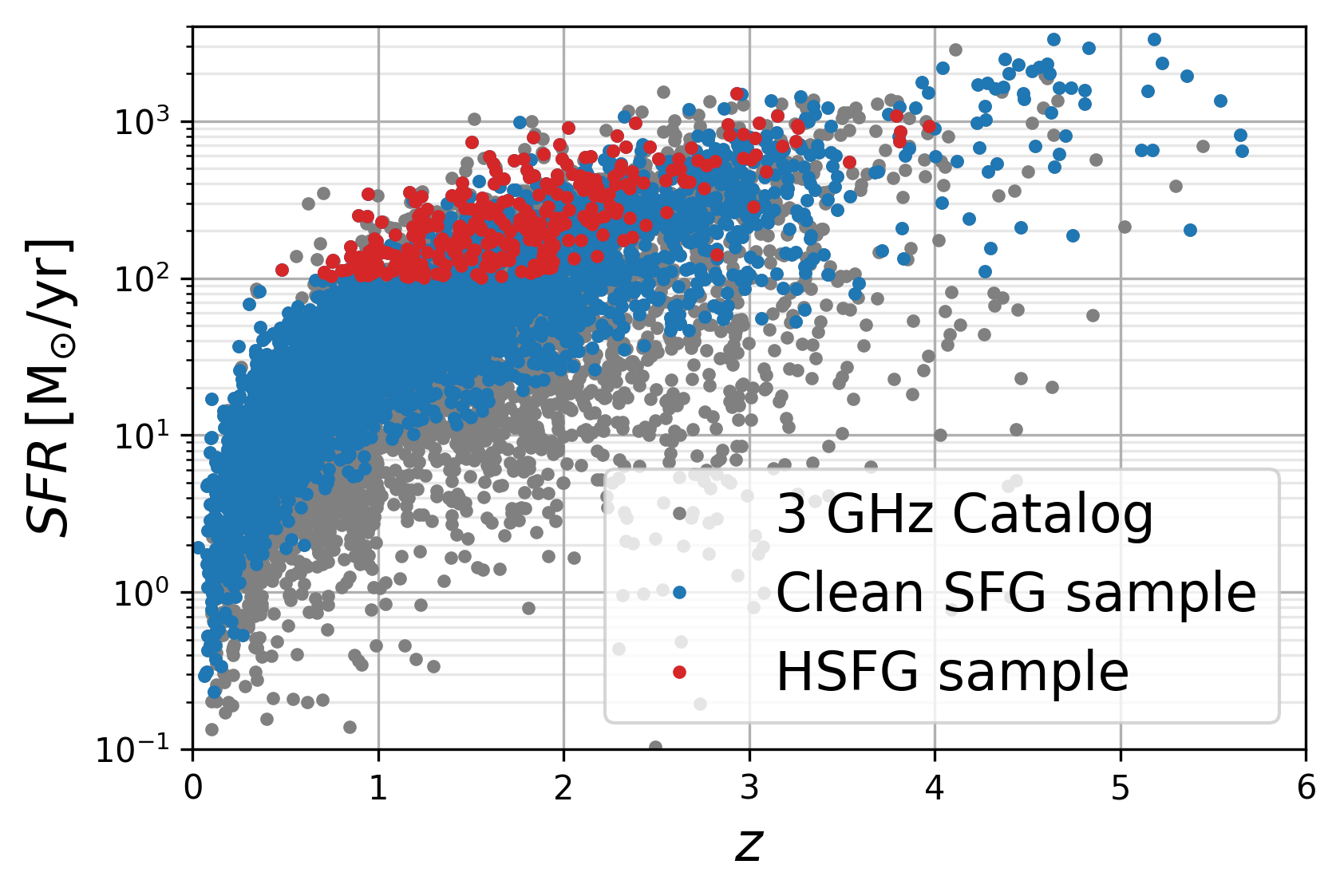}\hspace{10mm}\includegraphics[height=4cm]{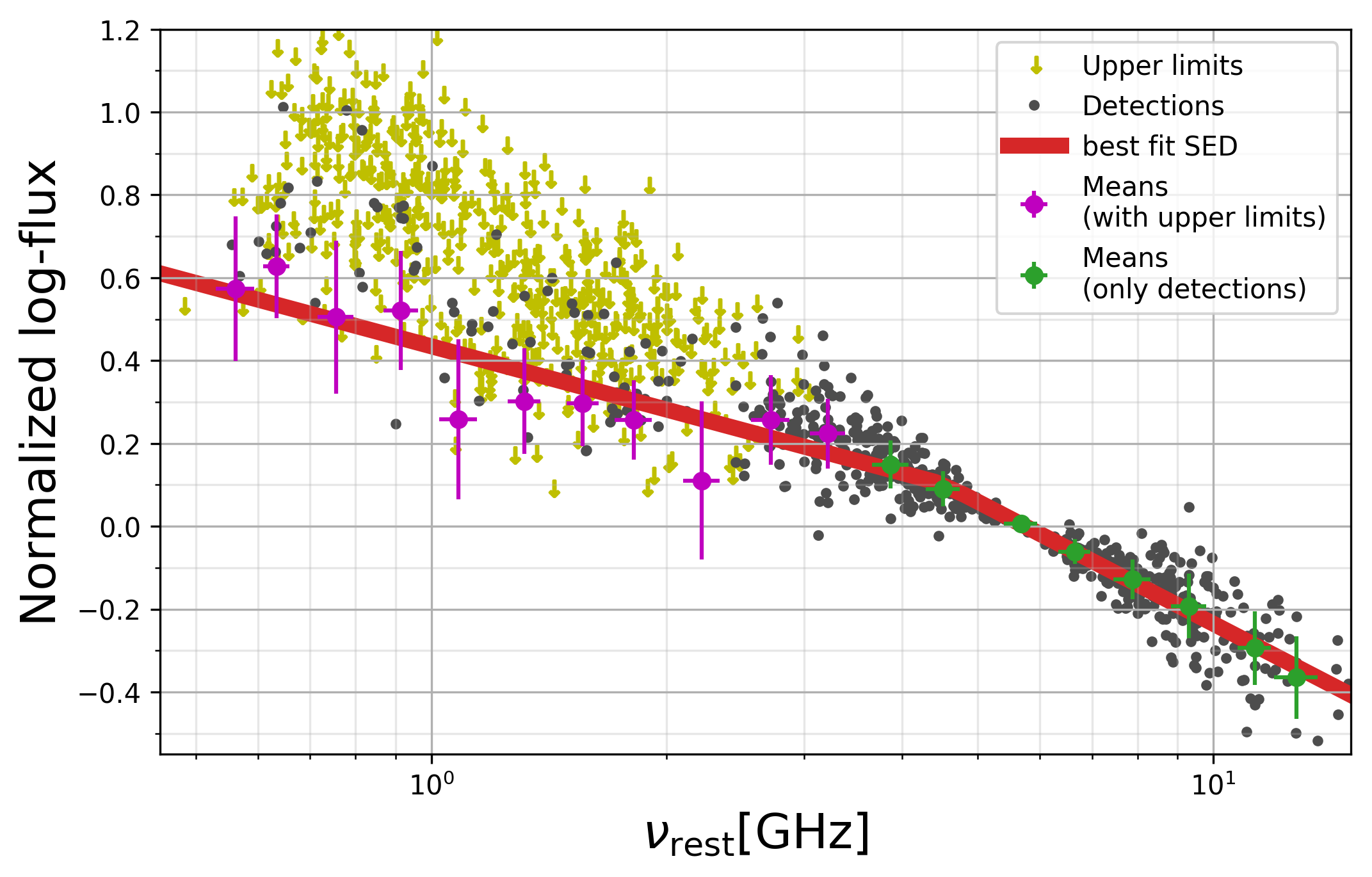}
\caption{The left panel shows SFRs versus redshift for the $3\GHz$ catalog and the HSFG sample, while the right panel shows the derived average radio SED for HSFGs.}\label{fig:1}
\end{figure}
Our sample is based on the VLA-COSMOS $1.4\GHz$  Joint Project \citep{Schinnerer07, Schinnerer10}, the  
VLA-COSMOS $3\GHz$ Large Project \citep{Smolcic17a} and previously unpublished $325\MHz$ and $610\MHz$ observations of the COSMOS field obtained with the Giant Metrewave Radio Telescope (GMRT).

The $325\MHz$ GMRT observations (project 07SCB01) towards the COSMOS field were conducted in 2007 over $45\hr$ in total. 
The data reduction and imaging were performed using the Source Peeling and Atmospheric Modeling pipeline (SPAM), as described in detail in \citet{Intema17}; see also Tisani\'c et al. (in prep.).
Observations at 610 MHz were conducted in 2011 over 19 pointings for 86 hours in total (project 11HRK01).
The data reduction, imaging and mosaicing were performed using the SPAM pipeline.
Primary beam and the average pointing error corrections were applied to the mosaic.
We used \emph{blobcat} \citep{Hales12}  to extract sources down to $5\sigma$, yielding 633 and 999 sources in the $325\MHz$ and $610\MHz$ maps, respectively.
The fluxes were corrected for bandwidth-smearing and resolved sources were determined, following  \citet{Bondi03}.
We start from the  ``clean star-forming galaxy sample", as defined in \citet{Smolcic17b}. This is the purest sample of star forming galaxies in the VLA-COSMOS 3 GHz Large Project  counterpart catalog \citep{Smolcic17b}, constructed by excluding active galactic nuclei (AGN) via a combination of criteria, such as X-ray luminosity, mid-infrared (color-color and SED based) criteria, $r^+-$NUV colors, and ($>3\sigma$) radio excess relative to the IR-based SFR, derived assuming the \citet{Chabrier03} initial mass function, as a function of redshift \citep[see Sect.~6.4 and Fig.~10 in ][]{Smolcic17b}. 
To determine physical properties of the galaxies, a three-component SED-fitting procedure including energy balance between UV-optical and IR radiation, as well as AGN contribution was applied using all  of the available photometry \citep[see][]{Delvecchio17}.
We select only objects from the ``clean star-forming galaxy sample" with IR-based SFRs greater than $100\Msun/\yr$ to minimize incompleteness above $z\approx 2$ (see left panel of Fig. \ref{fig:1}). We also require a $5\sigma$ detection at  both $1.4\GHz$ and $3\GHz$. 
We hereafter refer to this subset of 306 objects as the highly star-forming galaxy (HSFG) sample.
\section{Method}
To construct a typical SED for HSFGs we first normalize the spectra of individual galaxies at a particular rest-frame frequency via a power law fit to the observer-frame $1.4\GHz$ and $3\GHz$ data.  We then normalize the  $325\MHz$ and $610\MHz$ flux densities (using \emph{blobcat} measurements for detections, or $5\sigma$ upper limits, where $\sigma$ is the local RMS, for non-detections). Given that the redshifts of our sources range from $0.3$ to $4$, we constrain an average SED ranging over rest-frame frequencies from about $0.5$ to $15\GHz$.
Given that there are 531 upper limits and 693 data points in the measurements, we employ the survival analysis technique to derive the average SED of HSFGs over the whole redshift range, as described below. To achieve uniform frequency binning, we use 20 bins equally separated in log space of rest-frame frequencies. 
For each bin, in absence of upper limits, we compute the mean of the log-frequency, $\bar \nu$,  and its standard deviation, $\sigma_{\nu}$, the mean normalized log-flux density, $\bar F_\nu$, and its standard deviation, $\sigma_{F}$.
If there are upper-limits within a particular bin, we construct the Kaplan-Meier estimate of the survival function of fluxes within the bin, which we then fit with a Weibull distribution. We then use the parameters of the Weibull distribution to estimate $\bar F_\nu$ and $\sigma_{F}$.
We find that the standard combination of a steep power-law for synchrotron radiation and a flat power-law for free-free emission does not capture the flattening at low frequencies we observe in our SED.
Furthermore, we find no flattening at higher frequencies, which would have been indicative of the thermal contribution from the free-free emission.
We therefore decide upon using the broken power-law model for the log-normalized flux,  $\lg f$,
        	\begin{equation}
        	\lg f(\nu_{rest} | \alpha_1, \alpha_2, \nu_b, b)=\begin{cases}
            			-\alpha_2 \lg \frac{\nu_{rest}}{\nu_n} + b,& \nu_{rest}>\nu_b\\
                        -\alpha_1 \lg  \frac{\nu_{rest}}{\nu_n}+b+(\alpha_1-\alpha_2) \lg  \frac{\nu_{b}}{\nu_n},& \nu_{rest}<\nu_b
            		   \end{cases},
        	\end{equation}
where $\nu_n$ is the normalization frequency used to construct the normalized fluxes, $\nu_b$ the break frequency of the double power law, $\alpha_1$ ($
\alpha_2$) the spectral index below (above) $\nu_b$, and $\alpha_1, $ $\alpha_2$ and  $b$ are the free parameters of the model.
The fitting is performed using the Orthogonal distance regression method (hereafter, ODR), which accounts for the errors on both axes.
\section{Results}
 Due to the difficulty of fitting all of the parameters together, we fix the break frequency to the median value derived using the Markov Chain Monte Carlo (MCMC)  method, yielding $\nu_b=4.5\GHz,$ and fit the broken power law again using the ODR method.
 This procedure ensures optimal behavior of errors. The right panel of Fig. \ref{fig:1} shows our average radio SED. We find that the spectral index changes from $\alpha_1=0.51\pm 0.04$ below $\nu_b$ to $\alpha_2=0.98\pm0.07$ above $\nu_b.$
 
 The best fitting broken power law SED yields the $1.4\GHz$ rest-frame luminosities up to $20\%$ smaller relative to those derived using a simple, single-power law assumption with a spectral index of 0.7, with the largest offset occurring below $z\approx 1$.
 Above $z\approx 2$, the discrepancy diminishes as the $3\GHz$ observer-frame fluxes of galaxies above $z=2$ correspond to rest-frame fluxes above $9\GHz$. 
 This would in turn only shift the local $q$ values for highly star-forming galaxies towards larger values for $\sim0.1\,\mathrm{dex}$, while the $q$ values would remain similar for galaxies at higher redshifts.
We conclude that, assuming our SED holds for SFGs with $SFR>10\Msun/\yr$ and the mean \citet{Tabatabaei17} relation holds for SFRs  $<10\Msun/\yr$, we still observe a statistically significant decrease of $q$ with increasing redshift.

\section{Summary}
We  constrain the shape of the average radio SED for a sample of highly star-forming galaxies ($SFR>100\Msun/\yr$) in the COSMOS field based on the VLA $1.4\GHz$ and $3\GHz$ catalogs.
To constrain the radio SED below $1\GHz$, we have reduced the archival GMRT observations of the COSMOS field at $325$ and $610\MHz$ observer-frame frequencies. 
 
Accounting for non-detections in the GMRT maps using survival analysis and assuming a double power-law fit, we find that the power-law exponent changes from a spectral index of $\alpha_1=0.51\pm 0.04$ below a rest-frame frequency of $4.5\GHz$ to a steeper value of $\alpha_2=0.98\pm0.07$ above $4.5\GHz.$ 
Assuming our SED for SFGs with $SFR>10\Msun/\yr$ and the mean \citet{Tabatabaei17} relation for SFRs less than $10\Msun/\yr$,  we  observe a decrease of $q$ with increasing redshift. This suggests that the assumption of a  K-correction based on a simple power-law radio SED is not the likely cause of this trend, found in several previous studies.
\section*{Acknowledgements}
This research was funded by the European Union's Seventh Frame-work program under grant agreement 337595 (ERC Starting Grant, `CoSMass').


\begin{thebibliography}{}
\bibitem[Bondi et al.(2003)]{Bondi03} Bondi, M., Ciliegi, P., Zamorani, G., et al.\ 2003, \textit{A\&A}, 403, 857 
\bibitem[Chabrier(2003)]{Chabrier03} Chabrier, G.\ 2003, \textit{PASP}, 115, 763 
\bibitem[Calistro Rivera et al.(2017)]{CalistroRivera17} Calistro Rivera, G., Williams, W.~L., Hardcastle, M.~J., et al.\ 2017, \textit{MNRAS}, 469, 3468 
\bibitem[Clemens et al.(2008)]{Clemens08} Clemens, M.~S., Vega, O., Bressan, A., et al.\ 2008, \textit{A\&A}, 477, 95 
\bibitem[Condon(1992)]{Condon92} Condon, J.~J.\ 1992, \textit{ARA\&A}, 30, 575 
\bibitem[Delhaize et al.(2017)]{Delhaize17} Delhaize, J., Smol{\v c}i{\'c}, V., Delvecchio, I., et al.\ 2017, \textit{A\&A}, 602, A4 
\bibitem[Delvecchio et al.(2017)]{Delvecchio17} Delvecchio, I., Smol{\v c}i{\'c}, V., Zamorani, G., et al.\ 2017, \textit{A\&A}, 602, A3 
\bibitem[Galvin et al.(2017)]{Galvin17} Galvin, T., Seymour, N, Marvil, J, et al.\ 2017, \textit{arXiv}:1710.01967 
\bibitem[Hales et al.(2012)]{Hales12} Hales, C.~A., Murphy, T., Curran, J.~R., et al.\ 2012, \textit{MNRAS}, 425, 979 
\bibitem[Helou et al.(1985)]{Helou85} Helou, G., Soifer, B.~T., \& Rowan-Robinson, M.\ 1985, \textit{ApJ}, 298, L7 
\bibitem[Intema et al.(2017)]{Intema17} Intema, H.~T., Jagannathan, P., Mooley, K.~P., \& Frail, D.~A.\ 2017, \textit{A\&A}, 598, A78 
\bibitem[Ivison et al.(2010)]{Ivison10} Ivison, R.~J., Magnelli, B., Ibar, E., et al.\ 2010, \textit{A\&A}, 518, L31 
\bibitem[Kennicutt(1998)]{Kennicutt98} Kennicutt, R.~C., Jr.\ 1998, \textit{ARA\&A}, 36, 189 
\bibitem[Kennicutt(1998)]{Kennicutt98b} Kennicutt, R.~C., Jr.\ 1998, \textit{ApJ}, 498, 541  
\bibitem[Leroy et al.(2011)]{Leroy11} Leroy, A.~K., Evans, A.~S., Momjian, E., et al.\ 2011, \textit{ApJ}, 739, L25 
\bibitem[Magnelli et al.(2015)]{Magnelli15} Magnelli, B., Ivison, R.~J., Lutz, D., et al.\ 2015, \textit{A\&A}, 573, A45 
\bibitem[Schinnerer et al.(2007)]{Schinnerer07} Schinnerer, E., Smol{\v c}i{\'c}, V., Carilli, C.~L., et al.\ 2007, \textit{ApJS}, 172, 46 
\bibitem[Schinnerer et al.(2010)]{Schinnerer10} Schinnerer, E., Sargent, M.~T., Bondi, M., et al.\ 2010, \textit{ApJS}, 188, 384 
\bibitem[Scoville et al.(2007)]{Scoville07} Scoville, N., Aussel, H., Brusa, M., et al.\ 2007, \textit{ApJS}, 172, 1 
\bibitem[Smol{\v c}i{\'c} et al.(2017a)]{Smolcic17a} Smol{\v c}i{\'c}, V., Novak, M., Bondi, M., et al.\ 2017, \textit{A\&A}, 602, A1
\bibitem[Smol{\v c}i{\'c} et al.(2017b)]{Smolcic17b} Smol{\v c}i{\'c}, V., Delvecchio, I., Zamorani, G., et al.\ 2017, \textit{A\&A}, 602, A2 
\bibitem[Tabatabaei et al.(2017)]{Tabatabaei17} Tabatabaei, F.~S., Schinnerer, E., Krause, M., et al.\ 2017, \textit{ApJ}, 836, 185 
\bibitem[van der Kruit(1971)]{Kruit71} van der Kruit, P.~C.\ 1971, \textit{A\&A}, 15, 110 
\end{thebibliography}
\end{document}